\documentclass[11pt,twoside]{article}
\usepackage{asp2004}
\usepackage{psfig}
\usepackage{epsf}
\usepackage{graphics}
\usepackage{lscape}
\def\edcomment#1{\iffalse\marginpar{\raggedright\sl#1\/}\else\relax\fi}
\reversemarginpar

\begin{document}
\title{The Core Collapse Supernova Mechanism: Current Models, Gaps, and the Road Ahead}
\author{A. Mezzacappa}
\affil{Physics Division, Oak Ridge National Laboratory, Oak Ridge, TN 37831}

\begin{abstract}
The pursuit of the core collapse supernova explosion mechanism continues. While such efforts have been undertaken over the last four decades, it is only in the last decade that multidimensional models have been developed, and only in the last few years that significant components of supernova models, such as the neutrino transport, have been modeled with sufficient realism. We can now identify what are arguably the important components that must be included in any realistic three-dimensional supernova model. The challenge we now face is to include all of these in the models.
\end{abstract}
\thispagestyle{plain}

\section{Introduction}
Four hundred years after the observation of SN1604 by Galileo Galilei, the mechanism for core collapse supernova explosions remains unknown. Nonetheless, one-, two-, and three-dimensional simulations performed thus far have shown that (1) neutrino transport, (2) fluid instabilities, (3) rotation, and (4) magnetic fields, together with proper treatments of (5) the sub- and super- nuclear density stellar core equation of state, (6) the neutrino interactions, and (7) gravity will be important. This applies to the explosion mechanism  and to phenomena associated with supernova explosions, such as neutron star kicks, supernova neutrino and gravitational wave emission, and the polarization of supernova light. Not surprisingly, current two- and three-dimensional models have yet to include (1)--(4) with sufficient realism. One-dimensional spherically symmetric models have achieved a significant level of sophistication but, by definition, cannot incorporate (2)--(4), except phenomenologically. Fully general relativistic spherically symmetric simulations with Boltzmann neutrino transport do not yield explosions \citep{lmtmhb01}, demonstrating that some combination of (2), (3), and (4) is required to achieve this. (While the neutrino interactions and equation of state are important, we do not expect changes in them to lead to {\it qualitative} changes in the models.) It is important to note that layering the dimensionality and the physics will be needed to achieve a complete understanding of the supernova mechanism and phenomenology \citep{mlmhtb01}. The above mentioned one-dimensional simulations, for example, provided the first proof that (2), (3), and (4), in some combination, are {\it necessary} ingredients in the recipe for explosion.

The past modeling efforts alluded to above have illuminated that core collapse supernovae may be neutrino driven, MHD driven, or both \citep{s84,wm93,hbhfc94,bhf95,jm96,mw99,fw04,janka04}, but uncertainties in the current models prevent us from being able to single out any one of these possibilities. And it may be that more than one possibility is realized in Nature. Nonetheless, were a supernova neutrino driven, magnetic fields will likely have an impact on the dynamics of the explosion. Similarly, were a supernova MHD driven, the neutrino transport will dictate the dynamics of stellar core collapse, bounce, and the postbounce evolution, which in turn will create the environment in which the MHD-driven explosion would occur. Thus, while reduction will allow us to sort out the roles of each of the major physics components listed above, we will not obtain a quantitative, and perhaps even qualitative, understanding of core collapse supernovae until all components and their coupling are included in the models.

\section{Neutrino Transport}

In light of the above remarks, the ``neutrino heating mechanism," whereby the stalled supernova shock wave is reenergized by charged-current electron neutrino and antineutrino absorption on the dissociation-liberated protons and neutrons in the postshock flow, remains integral to the supernova paradigm, and neutrino transport is arguably the single most important component of any supernova model. Moreover, it has been shown [e.g., see \citet{jm96}] that the neutrino heating depends sensitively on the neutrino luminosities, spectra, and angular distributions in the ``gain'' (heating) region directly below the supernova shock wave, and these in turn are determined in the semitransparent region around the neutrinospheres, where neutrino transport is neither diffusive nor free streaming. In this region, only a solution to the neutrino Boltzmann kinetic equations can capture the three quantities of interest with sufficient accuracy. This is especially important when it comes to the neutrino spectra. The neutrino heating depends quadratically on the neutrino rms energies and only linearly on the neutrino luminosities and inverse flux factors (a measure of the isotropy of the neutrino distributions). With the exception of Wilson's spherically symmetric models that invoke the doubly diffusive neutron finger instability in the proto-neutron star to boost the neutrino luminosities, no simulation to date performed with multifrequency neutrino transport has yielded an explosion, a sobering fact \citep{wm93,sl94,mcbbgsu98b,rj00,bdm01,lmtmhb01,tbp03,burasetal03}. And this list now includes both one- and two-dimensional simulations. 
\footnote{A recent exception to this trend was found by \citet{janka04}, where a weak explosion of an 11 M$_{\odot}$ model was reported in the context of a two-dimensional model.}
Moreover, without neutron fingers, whose existence is a matter of current debate (see the discussion below), Wilson does not obtain explosions either \citep{w04}. 

As mentioned above, detailed spherically symmetric simulations that now include state of the art neutrino interactions, an industry standard equation of state, and multiangle, multifrequency, Boltzmann neutrino transport in full general relativity have been performed. Efforts by several groups are now underway to develop Boltzmann neutrino transport for the two- and three-dimensional cases \citep{cm03, cardall04,livneetal03}. The simulations performed thus far have been confined to the Newtonian gravity, $O(1)$ limit, restricted not only by Newtonian gravity but by the exclusion of the $O(v/c)$ ``observer corrections" (angular aberration and frequency shift) on the left-hand side of the Boltzmann equation. The observer corrections are critical in the evolution of the comoving frame neutrino distributions, and their inclusion in the models presents a significant technical challenge [e.g., see \citet{lmmbct04,cm03}], made increasingly difficult by the proliferation of terms in the Boltzmann equations as we move from one- to three-dimensional models and from Newtonian to general relativistic  models. Moreover, the memory and CPU requirements for three-dimensional Boltzmann neutrino transport are severe and will challenge any suite of algorithms and any computer architecture.

And it is now an experimental fact that neutrinos have mass and, therefore, mix in flavor. Observations of Solar and atmospheric neutrinos, and experiments at LSND, indicate there may be three independent values of the difference in the square of the neutrino masses ($\delta m^{2}$) and four mixing angles, which can be accommodated by including three active and one sterile neutrino in the models. Recent studies have shown that neutrino mixing in both active-active and active-sterile channels may occur deep in the stellar core even at small values of $\delta m^{2}$ and may alter the explosion dynamics, supernova nucleosynthesis, and terrestrial neutrino signatures, or all three \citep{qf95,fq04}. Neutrino mixing is a coherent, quantum mechanical phenomenon, unlike the incoherent collisional phenomena included in the Boltzmann kinetic equations. Therefore, a more complete (quantum kinetic) treatment of neutrino transport in stellar cores beyond (classical) Boltzmann transport is warranted.

\section{Fluid Instabilities}

The potential role of fluid instabilities in the post-stellar-core-bounce dynamics was first articulated  decades ago and explored phenomenologically for some time in the context of spherically symmetric models. Fortunately, two- and three-dimensional models that have emerged over the past decade have now allowed us to better explore the possibilities. 
(1) As has been pointed out by a number of authors [e.g., see \citet{hbhfc94,bhf95,burasetal03,fw04}], fluid instabilities have fundamentally altered the scenario under which neutrino shock reheating occurs. The existence of neutrino-driven convection below the supernova shock wave allows for both an explosive scenario, in which the shock wave and material behind it begin to move radially outward, and continued accretion, which maintains the neutrino luminosities sufficiently high so as to sustain the heating. In spherical symmetry, explosion and accretion are mutually exclusive. While earlier studies speculated that convection in the proto-neutron star might also be important, in boosting the neutrino luminosities, the most recent and complete studies indicate that convection in the proto-neutron star may not play a major role in the explosion [e.g., see \citet{burasetal03}]. This conclusion, of course, does not necessarily extend to other instabilities in the proto-neutron star, such as doubly diffusive instabilities, discussed below.
(2) Significant progress has been made on understanding the role of doubly diffusive instabilities in core collapse supernovae. In particular, detailed numerical experiments performed under conditions culled from core collapse supernova simulations suggest that neutron fingers are unlikely to occur in the stellar core early after bounce, during the shock reheating phase, and are therefore not expected to aid in generating an explosion \citep{brm04}.\footnote{It is important to note that when the Lattimer--Swesty equation of state \citep{ls91}, used in the above studies, is used by Wilson in his simulations, the criterion for neutron finger instability is not satisfied \citep{w04}. Whereas, when the Wilson equation of state is used, it is satisfied. This discrepancy must be explored.} However, these same experiments have led to the discovery of a new doubly diffusive instability (``lepto-entropy fingers") that may exist during this epoch and that may act, like neutron fingers, to boost the neutrinosphere luminosities. Future fully two- and three-dimensional models will be able to investigate this possibility.
(3) A completely new type of instability has recently been discovered in the core collapse supernova context \citep{bmd03} that may aid in generating the supernova and may be the underlying mechanism responsible for the generation of neutron star kicks and the polarization of supernova light. The postbounce stellar core flow is best characterized as an accretion flow through a quasi-stationary shock. It has been shown in two- and three-dimensional hydrodynamics studies constructed to reflect the conditions during the postbounce shock reheating epoch that nonspherical perturbations of the accretion shock lead to the development of a ``stationary accretion shock instability (SASI)." The SASI is an $l=1$ instability with a significant $l=2$ component, and results from the nonlinear feedback between vorticity introduced by the oblique supernova shock and nonspherical pressure waves generated by the interacting vortices trapped in the postshock flow that further distort the shock [this feedback mechanism was discussed in another context by \citet{foglizzo02}]. The potential ramifications of the SASI for the supernova mechanism and phenomenology were first elaborated in \citet{bmd03}. Recent studies \citep{jankaetal04} confirm the existence of this instability in models that include radial ray neutrino transport and explore in detail its consequences for neutron star kicks. It is important to note that, owing to the SASI, bipolar explosions can be produced even in the absence of rotation.

\section{Rotation and Magnetic Fields}

Neutrino transport in core collapse supernovae had been studied extensively over the last four decades, culminating recently in fully general relativistic simulations with Boltzmann neutrino transport that have essentially closed the book on spherically symmetric models [at least in the absence of neutrino mixing; \citet{lmtmhb01}]. And a number of two-dimensional simulations have now been performed over the past decade to explore the dynamics of fluid instabilities in the stellar core after bounce and their impact on the supernova mechanism, culminating in simulations that include sophisticated radial ray neutrino transport that  captures a significant amount of realism in the models \citep{burasetal03}. In contrast, very few simulations have been performed to date that include rotation, and no contemporary, sufficiently realistic simulations have been performed that include magnetic fields.

The work of several groups [e.g., see \citet{burasetal03,akiyamaetal03,fw04,tqb04}] has shown that rotation can significantly influence stellar core collapse and the postbounce dynamics in a variety of ways. (1) Centrifugal forces will slow collapse along the equator and, if sufficiently large, can lead to a low-density bounce. (2) Gravitational binding energy will be channeled differently during collapse relative to the spherically symmetric case. (3) The neutrinospheres will be distorted, and the neutrino luminosities and rms energies may be noticeably changed. (4) The preshock accretion ram pressure along the rotation axis and the equator may differ significantly. And rotation (5) will alter the development of fluid instabilities below the shock, (6) may provide a new source of internal energy in the postshock flow that may augment the energy supplied by neutrino heating, and (7) may have a dramatic impact on the growth of magnetic fields in the stellar core after bounce.

For example, the simulations by \citet{burasetal03} demonstrate clearly that rotation can have a dramatic effect on the postbounce shock dynamics. In their simulation, the average shock radius was increased by approximately a factor of 2 between 200 and 300 ms after bounce. Moreover, owing to centrifugal forces in the equatorial plane and a decreased ram pressure along the rotation axis, a violent overturn was observed in the postshock, convective region, and confined to an angular region around the rotation axis. These simulations were performed on a 90 degree angular grid, which will not admit the $l=1$ SASI instability mentioned above. Future simulations that include rotation and a 180 degree grid will allow us to more thoroughly explore whether postbounce instabilities and rotation act in concert in the explosion mechanism.

Again, the work of several groups [e.g., see \citet{s84,mpu02,akiyamaetal03,dq04,tqb04}] has shown that, much like rotation, magnetic fields can influence stellar core collapse and the supernova mechanism in a number of ways: (1) The development of significant magnetic pressure may have an impact on stellar core collapse and the postbounce flow. Magnetic fields may (2) alter the development of fluid instabilities in the stellar core after bounce, (3) provide additional channels for the generation of internal energy through viscous dissipation, (4) alter the weak interactions in the stellar core, and (5) supply a significant, and perhaps dominant, fraction of the luminosity powering the explosion. Arguably, the fundamental question is whether the magnetic fields will organize into configurations that will drive and collimate outflows from the stellar core. 

The pioneering simulations of \citet{lw70} and \citet{s84} were the first to explore the evolution of stellar core magnetic fields during core collapse and their impact on the explosion mechanism. These simulations exhibited the development of a magnetic bubble deep in the core owing to the dramatic increase in magnetic pressure close to the rotation axis as core field lines are dragged inward and compressed. This magnetic bubble led to buoyant, bipolar outflows that culminated in bipolar explosions. However, \citet{s84} concluded that inordinately large rotation and magnetic fields strengths were required for an explosion to develop. The magnetic fields in the stellar core can be amplified in the way just described, but the growth of field strength through other mechanisms must also be considered. In particular, at the time of the early simulations by Leblanc, Wilson, and Symbalisty, the magnetorotational instability (MRI) was not yet discovered. The MRI was first discovered in the context of the differential rotation in accretion disks \citep{bh91} but may operate in a differentially rotating core after stellar core bounce. \citet{wmw02} delineated the possible roles organized magnetic fields could play in the supernova mechanism, and \citet{akiyamaetal03} were the first to propose that the MRI could be important in the core collapse supernova context. They argued that, with sufficient differential rotation, the MRI could amplify the magnetic field strengths in the stellar core exponentially quickly and lead to magnetohydrodynamic luminosities $\sim 10^{52}$ erg/s, rivaling the neutrino luminosities. 

However, it is important to note that improved progenitor models that include both rotation and magnetic breaking yield iron cores that rotate at only $\sim 0.1$ rad/s \citep{hws04}. Without magnetic breaking, the core rotation rates are 1--2 orders of magnitude faster. Therefore, magnetic fields lead to competing outcomes: For the MRI to increase field strengths to magnitudes that would be dynamically significant, perhaps dominant, would require significant differential rotation in the core. On the other hand, magnetic breaking in the core tends to slow the stellar core rotation prior to collapse. And, of course, two- and three-dimensional progenitor models will be required to better determine the differential rotation of the core at the start of collapse.

Magnetic fields will likely play an important role in supernova dynamics. Whether they will take center stage as the main source of luminosity powering the explosion remains to be seen, and generally, their precise role will require realistic three-dimensional simulations. None have yet been performed.

\section{Neutrino Interactions, Sub- and Super-Nuclear Density EOS}

Improvements in modeling the macrophysics of stellar core radiation magnetohydrodynamics must be matched by improvements in modeling the neutrino weak interactions in the stellar core and the stellar core equation of state. 

Recent simulations \citep{hixetal03} have demonstrated that electron capture during stellar core collapse is dominated by capture on nuclei, not protons, and that a more sophisticated treatment of capture leads to dynamically significant changes in the size of the inner homologous core (and therefore, where the shock is launched) and the preshock stellar core profiles through which the shock will move, which will affect the subsequent shock dynamics, development of core fluid instabilities, and nucleosynthesis.  Significant progress has been made in the development of nuclear structure models for heavy nuclei, but a substantial effort will be required to extend these models to the nuclei well above mass 100 that populate the stellar core during collapse.

The naive picture of neutrinos interacting on free nucleons and nuclei has now been replaced by the more accurate picture of neutrinos interacting with correlated nuclei and nucleons at sub- and super-nuclear density. Calculations of neutrino interactions on nuclei during stellar core collapse that include correlations among nuclei, neutrino interactions on the extended structures present in the nuclear ''pasta" phase during the transition from nuclei to nuclear matter, and neutrino interactions on strongly interacting, correlated nucleons in the proto-neutron star have been completed [for a review, see \citet{brt04}]. Nonetheless, much more remains to be calculated, and these calculations will be extremely challenging. For example, in the case of neutrino interactions in nuclear matter, pioneering efforts have thus far invoked the random phase approximation (RPA), which includes some but not all of the correlations present in the high-density proto-neutron star medium.

Simulations have been performed recently that explored the sensitivity of supernova models to variations in the stellar core equation of state (EOS). Core collapse and postbounce simulations using the Lattimer--Swesty compressible liquid droplet EOS \citep{ls91}, the Shen et al. relativistic mean field EOS \citep{shenetal98}, and the Hillebrandt--Wolf Hartee--Fock EOS \citep{hw85} yield, for example, three-flavor neutrino luminosities that differ by tens of percent between 50 and 150 ms after bounce \citep{janka04}, which is quantitatively significant. Therefore, further development of the equation of state, and such comparisons, must continue.

\section{Gravity}

One should expect the gravitational fields around proto-neutron stars to deviate significantly from their Newtonian values. In fact, detailed comparisons of Newtonian and fully general relativistic collapse have been performed in spherical symmetry by \citet{bdm01} and confirm this expectation. The results from models beginning with 15 and 25 M$_{\odot}$ progenitors show that (1) the neutrinosphere, gain, and shock radii and (2) the infall velocities can differ by as much as a factor of 2 between the Newtonian and general relativistic models. In light of these results, Newtonian multidimensional models can only be viewed as a stepping stone to fully general relativistic, realistic multidimensional models. 

\section{Conclusion}

Much work remains to determine the core collapse supernova explosion mechanism and to understand all of the phenomena associated with these explosions. Despite significant progress and the accumulation of a wealth of understanding over the last four decades, as well as a number of specific clues, we still cannot answer some of the most fundamental questions. And, certainly, multidimensional supernova modeling is only in its infancy. However, guiding observations are becoming increasingly sophisticated and supercomputers are becoming increasingly powerful. Consequently, we are now presented with a unique opportunity to respond to this challenge.

\acknowledgments{
A.M. is supported at the Oak Ridge National Laboratory, which is 
managed by UT-Battelle, LLC for the DOE under contract DE-AC05-00OR22725.
He is also supported in part by a DOE ONP Scientific Discovery through 
Advanced Computing Program grant.}

\end{document}